\documentclass[aps,prb,twocolumn,groupedaddress]{revtex4}
\usepackage{amsmath}
\usepackage{amssymb}
\usepackage{graphicx}

\begin{document}



\title{Spin-Orbit Berry Phase in a Quantum Loop}


\author{Maxim P. Trushin and  Alexander L. Chudnovskiy}
\affiliation{1. Institut f\"ur Theoretische Physik, Universit\"at Hamburg,
Jungiusstr 9, D-20355 Hamburg, Germany}


\date{\today}

\begin{abstract}
We have found a manifestation of spin-orbit Berry phase in the conductance 
of a mesoscopic loop with Rashba spin-orbit coupling placed in an external magnetic field
perpendicular to the loop plane.
In detail, the transmission probabilities for a straight quantum wire 
and for a quantum loop made of the same wire have been calculated and 
compared with each other.
The difference between them has been investigated and identified with
a manifestation of spin-orbit Berry phase.
The non-adiabaticity effects at small radii of the loop have been found as well.
\end{abstract}

\pacs{73.23.Ad \sep 05.60.Gg \sep 03.65.Vf}
\keywords{ballistic transport \sep spin-orbit coupling \sep Berry phase}

\maketitle

\section{Introduction}

\label{intro}

The beauty of the topological Berry phase concept 
\cite{ProcR1984berry} inspires much
theoretical and experimental activity aimed at finding
its manifestations in different areas of modern physics
\cite{shapere1989}. Berry describes a quantal 
system in an eigenstate, slowly transported around a 
closed path in the phase space by varying parameters in its Hamiltonian.
According to the adiabaticity theorem, if the Hamiltonian
is returned to its original form, the system will
return to its original state, apart from a phase factor.
In addition to the familiar dynamical phase,
such a state can acquire a geometrical, path-dependent
phase factor, which is the result of the adiabatic variation
of the external parameters. This phase is known as Berry's phase.
(See also a fundamental generalization of this idea
for a non-adiabatic evolution \cite{PRL1987aharonov}.)

A possible candidate for
the role of such an external parameter in solid state physics
is the external magnetic field
${\mathbf B}$ that interacts with the electron spin via the Zeeman effect.
This interaction is described by the following Hamiltonian
\begin{equation}
\label{zeeman}
H_Z=\frac{g\mu_B}{2}{\bf \sigma \cdot B},
\end{equation} 
where ${\bf \sigma}=\{\sigma_x,\sigma_y,\sigma_z\}$
are the Pauli matrices, and $\mu_B$, $g$ are Bohr magneton
and g-factor respectively.
When the value of the magnetic field is constant and its
direction follows adiabatically a closed trajectory, the
spin wave function acquires the topological phase 
which is proportional to the solid angle
subtended in a space by the magnetic field \cite{ProcR1984berry}.

The possibility to control the Berry phase by means of the Zeeman effect
is the central issue explored in the pioneering 
\cite{PRL1990loss,PRB1992loss,PRL1992stern} and recent \cite{PRL1998lyanda,PRB2000engel,PRL2001frustaglia,PRB2004frustaglia,PRB2003popp,PRB2004hentschel1,PRB2004hentschel2} papers.
In particular, the authors consider
the adiabatic  as well as non-adiabatic motion of electrons through a mesoscopic ring
in the presence of a static, inhomogeneous magnetic field.
It is shown that the Berry phase, accumulated by the 
spins of electrons encircling the ring,
leads to persistent equilibrium charge and spin currents
\cite{PRL1990loss,PRB1992loss} or 
affects the conductance of the ring 
\cite{PRL1992stern,PRB2000engel,PRB2004hentschel1} in a way similar to the Aharonov-Bohm effect \cite{PR1959aharonov}.
The latter point is of particular interest to the topic.
Indeed, since Aharonov-Bohm and Berry phases
can be varied individually, the
interplay of the two phases yields a rich variety of behavior. In particular,
the amplitudes of the Aharonov-Bohm oscillations are strongly affected by the Berry phase \cite{PRB2000engel}.
Moreover, the authors of Ref.~\cite{PRB2000engel} 
show that these amplitudes can be completely suppressed at certain magic
tilt angles of the external fields. 


As was noted above, in order to observe the geometric phase in an electronic system with spin, the application of an orientationally inhomogeneous (e. g. radial)
magnetic field is necessary.
However, the manner in which the magnetic field is varied in Refs.~\cite{PRL1990loss,PRB1992loss,PRL1992stern,PRL1998lyanda,PRB2000engel,PRL2001frustaglia,PRB2004frustaglia,PRB2003popp,PRB2004hentschel1,PRB2004hentschel2}
leads to rather difficult experiments.
Fortunately, the desired magnetic field texture can be experimentally implemented via fabricating 
the loop (or ring) from a material with spin-orbit interactions 
of Rashba type \cite{JETPL1984bychkov}.
Indeed, the Rashba operator $H_R=\alpha\left[{\bf \sigma}\times{\bf k}\right]_z$
can be rewritten for the loop of radius $R$
in the quasi-classical limit ($kR\gg 1$) as
\begin{equation}
\label{rashba}
H_R=\alpha\left(\sigma_x\cos\varphi+
\sigma_y\sin\varphi\right)\left(-\frac{i}{R}\frac{\partial}
{\partial\varphi}\right).
\end{equation}
(Here, $\alpha$ is the Rashba constant.)
The effect of Rashba spin-orbit coupling on the electron motion
in the ring is seen clearly from Eq.~(\ref{rashba}): namely,
the electrons in such a ring 
experience a radial built-in Zeeman-like magnetic field
\begin{equation}
\label{Bin}
B_{\mathrm{in}}=\frac{2\alpha k}{g\mu_B},
\end{equation}
where $k$ is the characteristic wave vector.
In other words, the Rashba effect in the quasiclassical limit represents
the effective Zeeman-like magnetic field $B_{\mathrm{in}}$.
It is important to emphasize, that this in-plane magnetic field
{\em does not} relate to the real external
magnetic field $B_{\mathrm{ext}}$, but stems 
from the internal properties of the substance (spin-orbit interactions).
Most important, however, the external $B_{\mathrm{ext}}$ and in-plane $B_{\mathrm{in}}$ components
form the desired inhomogeneous magnetic field texture
and in that way can provide the geometric phase indications
through interference patterns in the conductance of the ring.
This pretty idea is attracting both theoretical
\cite{PRL1993aronov,PRB1994zhou,PRL1993lyanda,PRL1994qian,PRB1999mal'shukov}
and  experimental
\cite{PRL1998morpurgo,MEE1999nitta,PRL2002yau,physE2004yang,EPL2004yang} attention.

Let us consider the geometric phase acquired by the wave function of a charged {\em and}
spin-full particle as it travels around the ring structure with Rashba spin-orbit coupling.
The system is placed in the
external magnetic field $B_{\mathrm{ext}}$, which is perpendicular to the ring
plane.
Firstly, since the particle carries a charge, it picks up an Aharonov-Bohm phase \cite{PR1959aharonov}
\begin{equation}
\label{AB}
\phi_{\mathrm{AB}}=2\pi\frac{\Phi}{\Phi_0},
\end{equation}
where $\Phi_0$ is the flux quantum, and $\Phi=\pi R^2 B_{\mathrm{ext}}$ is the magnetic flux enclosed by the ring. Secondly, if the particle carries a spin of $1/2$ and its motion is adiabatic, then the spin geometric phase, according to Berry's definition \cite{ProcR1984berry}, reads
\begin{equation}
\label{berry}
\phi_B=
\pi\left(1-\frac{B_{\mathrm{ext}}}{\sqrt{B^2_{\mathrm{ext}}+B^2_{\mathrm{in}}}}\right),
\end{equation}
and the full geometric phase is a sum of both 
$\phi_B$ and $\phi_{\mathrm AB}$.
Note that the adiabaticity requires comparatively large values of $B_{\mathrm{in}}$ and $B_{\mathrm{ext}}$ so that the electron spin precesses few times within a cycle.

In Ref.~\cite{PRL1993aronov}, the authors established a
one-particle Hamiltonian for electrons moving on a 1D ring in the presence of Rashba spin-orbit coupling
and Zeeman splitting.
Furthermore, the ballistic motion of electrons in the
absence of scattering and spin-flip processes has been studied.
In the spirit of the seminal paper
by B\"uttiker, Imry and Azbel 
\cite{PRA1984buettiker},
the transmission amplitude of the ring has been derived and the conductance oscillations have been investigated.
We should note, however, that authors of Ref.~\cite{PRL1993aronov} used a non-Hermitian operator 
in the Hamiltonian. Zhou, Li, and Xue \cite{PRB1994zhou} noticed this
fact and derived a different (Hermitian) Hamiltonian operator. However, in their
Hamiltonian the spin-orbit coupling originates from an electric field pointing in the radial
direction and not in the direction perpendicular to the plane of the ring.
This is not the correct Rashba term for inversion layers \cite{JphC1984bychkov}. 
The procedure for obtaining the correct Hamiltonian
has been described in Ref.~\cite{PRB2002meijer}.

In spite of the mentioned shortcoming, Ref.~\cite{PRL1993aronov} has been
the stimulus for the subsequent studies.
In particular,
topological transitions in the ring conductance interference
pattern subject to Berry's phase have been studied in \cite{PRL1993lyanda}.
It manifests itself in a steplike conductance-magnetic field and
conductance-gate voltage characteristics.
The transition takes place when the Berry phase is dropped by an additional
static magnetic field $B_{\mathrm{ext}}$ from odd of $\pi$
to zero as it follows from Eq.~(\ref{berry}).
The non-adiabatic spin-orbit geometric phase
(of non-Berry, but Aharonov-Anandan type \cite{PRL1987aharonov}) in quantum rings 
has been investigated in Ref.~\cite{PRL1994qian}.
It has been shown that such a 
phase $\phi_\mathrm{AA}$ becomes the spin-orbit Berry phase
$\phi_B$ in the adiabatic limit.
In order to analyse the structure of the Aharonov-Bohm oscillations influenced
by the spin-orbit Berry phase,
the Fourier spectra of conductance in a two-dimensional ring have been 
calculated \cite{PRB1999mal'shukov}. 
Note that the method of Fourier analysis is the only
suited one for comparison of the theoretical results with the experimental data discussed below.

Another important feature of electron transport through the ring with
the Rashba coupling is that, even in the absence of an external magnetic
field, the topological effects due to the Aharonov-Casher phase 
$\phi_\mathrm{AC}$ \cite{PRL1984aharonov} can
take place. (The discussion of relations between $\phi_\mathrm{AA}$, $\phi_\mathrm{AC}$,
and $\phi_B$ can be found in the section \ref{sec3}.)
There is a recent series of articles 
\cite{APL2004molnar,PRB2005molnar,PRB2005foeldi} which deals
with spin-dependent transmission through one-dimensional quantum rings subject to 
the Aharonov-Casher phase. Moreover, ballistic electron transport
through {\em chains} of rings is studied.
However, in contrast to our approach, the Zeeman effect is neglected. 

In pioneering observations of Berry phase \cite{PRL1998morpurgo,MEE1999nitta},
the Aharonov-Bohm oscillations were studied in InAs two-dimensional two-contact quantum rings with strong
spin-orbit interaction. The Fourier transforms of over 30 traces of oscillations were averaged and a small splitting of the main peak in the final Fourier spectrum was interpreted as a possible manifestation of the spin Berry phase.

An attempt has been made to observe Berry phase in
quantum rings fabricated in a GaAs/AlGaAs heterostructure with a 2D hole system \cite{PRL2002yau}.
In such a setup, the inversion asymmetry results from the 
GaAs zinc blende crystal structure as well as from an electric field, which is perpendicular to the 2D plane.
Along with the main peak whose frequency corresponds to the magnetic flux enclosed by the ring,
there are some extra peaks in the Fourier spectra of the measured Aharonov-Bohm oscillations.
A qualitative comparison of the Fourier transforms with its simple simulation provides a striking demonstration of the Berry phase.

\begin{figure}

  \includegraphics{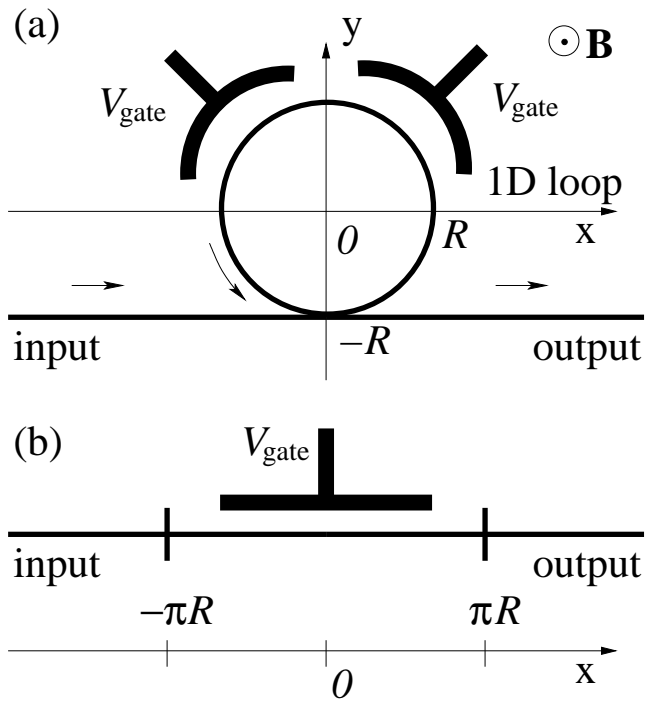}

\caption{Geometry of the systems under consideration.
(a) Quantum loop of radius $R$. Note, that in contrast to Ref.~\cite{EPL2004yang}
the electron beam {\em does not} split while it enters the loop.
(b) Quantum wire of the length $2\pi R$ which is made of the {\em same}
material as the quantum loop.}
\label{fig1}       
\end{figure}

\begin{figure*}

  \includegraphics{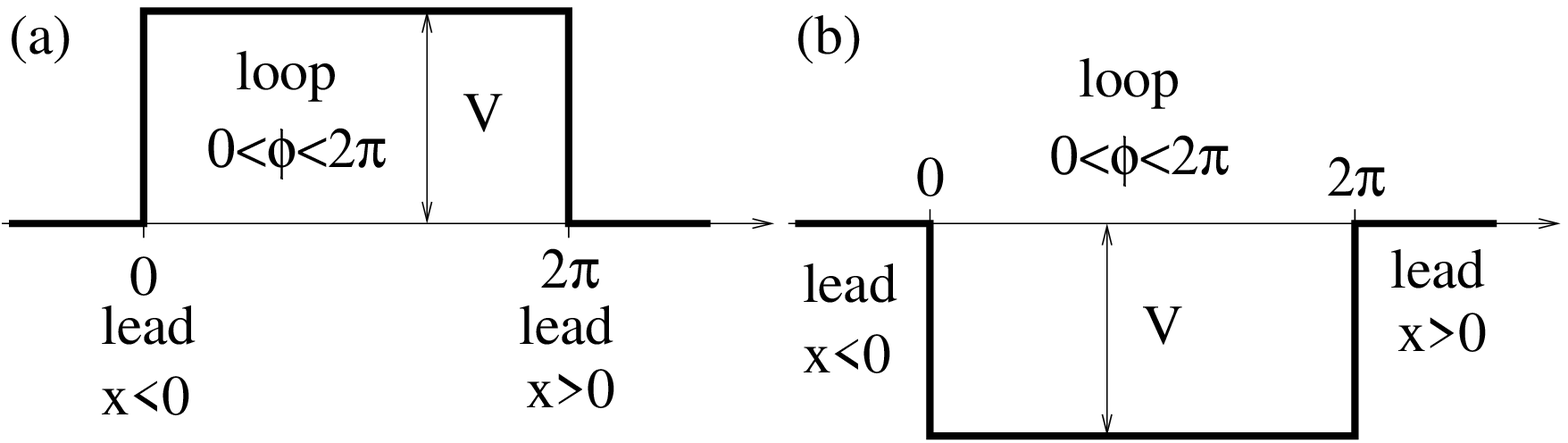}

\caption{Two variants of the potential profile adopted in the solution.
The bottoms of the bands can be lifted (a) 
or pulled down (b) in the loop region by $V$.}
\label{ch1fig1a}       
\end{figure*}

In contrast to earlier work, the authors of Refs.~\cite{physE2004yang,EPL2004yang} 
furnish a novel configuration, in which the ballistic ring
forms {\em one} collimating contact with the tangential current lead.
Beside the absence of unknown asymmetry in the arm length
(that always gave an uncertainty in a two-contact configuration) and additional spin rotation at contacts,
such a setup allows to let
only one transverse mode with a small longitudinal momentum enter the ring through the contact. Such a setup allows direct observations of the spin-orbit
Berry phase in conductance oscillations.

Finally, we would like to notice the recent paper \cite{condmat2005koenig},
where a manifestation of the geometric Aharonov-Casher phase 
is found in the conductance
of quantum rings fabricated from HgTe/HgCdTe quantum wells.
In these structures, the Aharonov-Bohm oscillations 
exhibit significant changes when the Rashba coupling is
varied by means of the gate voltage.
It is shown that these changes are due to the geometric phase contribution
to the electronic wave function.

In present paper, we study theoretically the system similar to \cite{physE2004yang,EPL2004yang}.
There are, however, some important differences.
First, the possibility for electrons to bypass the ring is assumed to be negligible in our system.
Therefore, the electron beam does not split while it enters or leaves the ring.
Thus, we study rather {\em a quantum loop} (Fig.~\ref{fig1}a)
than {\em a quantum ring} connected to the tangential lead
\cite{physE2004yang,EPL2004yang}. That is why, the Aharonov-Bohm effect does not take place here.
Second, we consider the modulation of the potential profile in the loop region by
means of the gate voltage applied to the structure. Although the realization of such
a setup requires rather complicated design,
the gated InAs rings have been fabricated \cite{priv}.
And finally, our system is purely one-dimensional, while this is not the case in 
Refs.~\cite{physE2004yang,EPL2004yang}.
However, so long as the gate voltage can be applied,
the upper size quantization subbands can be easily depopulated so, that
only a single band is occupied. Therefore,
the one-dimensionality of the quantum loop is not a big problem anymore.
Thus, the collimating contacts provided in Refs.~\cite{physE2004yang,EPL2004yang}
could be a powerfull tool for topological phase investigations
in (quasi-)one-dimensional systems as one studied below.

The detailed description and solution of the model as well as 
the discussion of results obtained is given in the next sections.

\section{Model}

\label{sec1}

In order to find the transmission probability through the quantum loop,
we have to solve the corresponding Schr\"odinger equation.
To this end we divide the system in three parts: input channel, the loop itself (which is 
actually the arc of $2\pi$-length) and output channel.
The Hamiltonians describing the propagation of electrons in the input/output channels read
\begin{equation}
\label{1Dwire_hamiltonian}
H_\mathrm{wire}=\left( \begin{array}{cc}
\frac{\hbar^2}{2m^*}\hat{k}_x^2 + \varepsilon_Z &
i\alpha\, \hat{k}_x \\
-i\alpha\, \hat{k}_x  &
\frac{\hbar^2}{2m^*}\hat{k}_x^2 - \varepsilon_Z \end{array} \right),
\end{equation}
whereas the propagation through the loop of radius $R$ is governed by the Hamiltonian
\begin{equation}
\label{loop_hamiltonian}
H_\mathrm{loop}=\left( \begin{array}{cc}
\varepsilon_0\,\hat{q}_{\varphi}^2 + \varepsilon_Z + V &
\frac{\alpha}{R}\,{\rm e}^{-i\varphi}\left(\hat{q}_\varphi - \frac{1}{2}\right) \\
\frac{\alpha}{R}\,{\rm e}^{i\varphi}\left(\hat{q}_\varphi + \frac{1}{2}\right)  &
\varepsilon_0\,\hat{q}_{\varphi}^2 - \varepsilon_Z  + V \end{array} \right).
\end{equation}
Here $\hat{k}_x=-i\frac{\partial}{\partial x}-\frac{\Phi}{\Phi_0}\frac{1}{R}$,
$\hat{q}_\varphi=-i\frac{\partial}{\partial\varphi}-\frac{\Phi}{\Phi_0}$
are momentum and angular momentum operators respectively,
$\Phi=\pi R^2 B_z$ is the magnetic flux, $\Phi_0$ is the flux quantum,
$\varepsilon_0=\hbar^2/(2m^*R^2)$ is the size confinement energy
with the effective electron mass $m^*$,
$\varepsilon_Z=g^*\mu_B B_z/2$ is the Zeeman energy,
and $V$ denotes the energy shift determined by the
gate voltage applied to the loop. (See Fig.~\ref{ch1fig1a} for the 
examples of the profile studied below.)

We adopt the vector potential $\mathbf{A}$ to be tangential to the direction of the current.
Thus, in the loop we choose $\mathbf{A}(x,y)=\frac{1}{2}B_z\left(x\,\mathbf{j}-y\,\mathbf{i}\right)$,
or, in cylindrical coordinates, $A_\varphi (\varphi)=\Phi/2\pi R$, whereas
the vector potential in the input and output channels is determined by the continuity condition 
at the junction point with the loop itself ($x=0$, $y= -R$); hence we have $A_x=\Phi/2\pi R$.

We denote the wave functions for each part as $\Psi_\mathrm{loop}^\pm(\varphi)$ for the loop,
$\Psi_\mathrm{in}^\pm(x)$ and $\Psi_\mathrm{out}^\pm(x)$ for input and output
channels respectively.
In order to find the wave function describing the whole system,
we impose the boundary conditions that warrant the continuity of the wave function and
its first derivative at the boundaries between the loop and input/output channels
\begin{equation}
\label{conditions}
\left\{\begin{array}{l}
\left(\Psi_\mathrm{in}^+ + \Psi_\mathrm{in}^-\right)|_{x=0}
= \left(\Psi_\mathrm{loop}^+ + \Psi_\mathrm{loop}^-\right)|_{\varphi=-\pi/2}, \\
\left(\Psi_\mathrm{loop}^+ + \Psi_\mathrm{loop}^-\right)|_{\varphi=3\pi/2}=
\left(\Psi_\mathrm{out}^+ + \Psi_\mathrm{out}^-\right)|_{x=0}, \\
\left(\nabla\Psi_\mathrm{in}^+ + \nabla\Psi_\mathrm{in}^-\right)|_{x=0}
= \left(\nabla\Psi_\mathrm{loop}^+ + \nabla\Psi_\mathrm{loop}^-\right)|_{\varphi=-\pi/2}, \\
\left(\nabla\Psi_\mathrm{loop}^+ + \nabla\Psi_\mathrm{loop}^-\right)|_{\varphi=3\pi/2}=
\left(\nabla\Psi_\mathrm{out}^+ + \nabla\Psi_\mathrm{out}^-\right)|_{x=0}.
\end{array}\right.
\end{equation}
The operator $\nabla$ is given by
$\nabla=\frac{1}{R}\frac{d}{d\varphi}$ in the loop region, and $\nabla=\frac{d}{d x}$
in the input and output channels.

In the next section, we find the electron eigen states for the loop, the input, 
and the output channel,
and solve the system of equations (\ref{conditions}).
The solution gives us the transmission and reflection amplitudes
(and, as consequence, the transmission/reflection probabilities)
for each spin mode.

\section{Solution of the problem}
\label{sec2}

Let us start from the input channel. The Hamiltonian (\ref{1Dwire_hamiltonian})
acts in SU(2) spin space. The corresponding Schr\"odinger equation allows two solutions
\begin{equation}
\label{input_psi+}
\Psi^+_\mathrm{in}(x)={\mathrm e}^{\frac{i \Phi }{\Phi_0 R}x}
\left(\begin{array}{c}
\cos\gamma^+\left({\mathrm e}^{i k^+ x} + A^+ {\mathrm e}^{-i k^+ x}\right) \\
-i\sin\gamma^+\left({\mathrm e}^{i k^+ x} - A^+ {\mathrm e}^{-i k^+ x}\right) \end{array} \right),
\end{equation}
\begin{equation}
\label{input_psi-}
\Psi^-_\mathrm{in}(x)={\mathrm e}^{\frac{i \Phi }{\Phi_0 R}x}
\left( \begin{array}{c}
-i\sin\gamma^-\left({\mathrm e}^{ i k^- x} - A^- {\mathrm e}^{- i k^- x}\right) \\
\cos\gamma^-\left({\mathrm e}^{ i k^- x} + A^- {\mathrm e}^{-i k^- x}\right) \end{array} \right),
\end{equation}
where
\begin{equation}
\tan\gamma^\pm= -\frac{{\varepsilon_Z}}{{k^{\pm}}\,{\alpha}}  +
\sqrt{1 + \left(\frac{{\varepsilon_Z}}{{k^{\pm}}\,{\alpha}}\right)^2},
\end{equation}
and ``$\pm$'' are the spin indices.

\begin{figure*}
\includegraphics{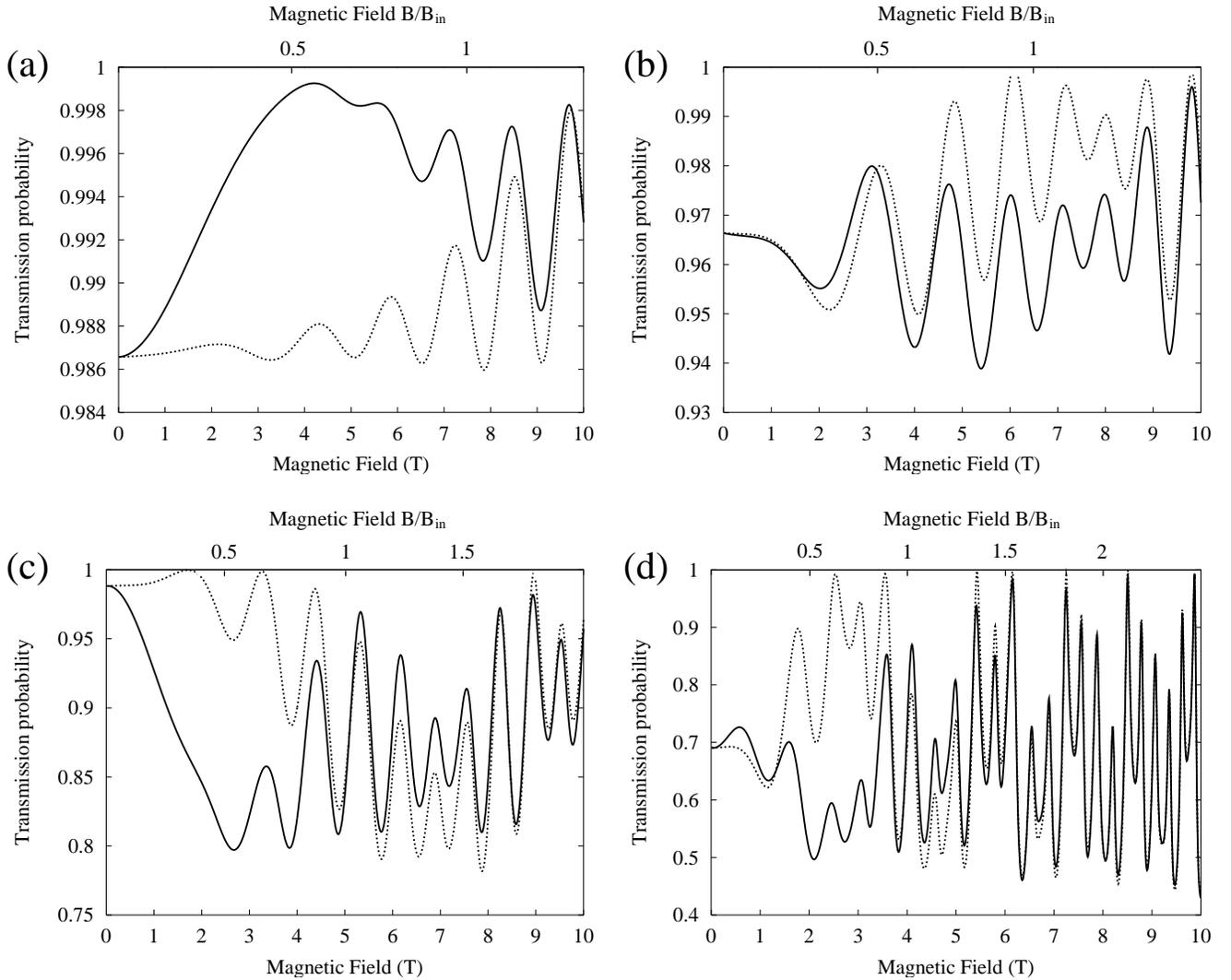}
\caption{Transmission probabilities
for the loop of radius $R=5\cdot 10^{-5}$cm
(solid lines) and corresponding straight wire of length
$L=2\pi R$ (dotted lines) versus external magnetic field.
Magnetic field is given in Tesla (lower axis) as well as
in units of $B_\mathrm{in}$ (upper axis).
Each panel corresponds to different height of the barrier $V$:
(a) $V=6.25$meV ($V/E_F\approx 0.2$),
(b) $V=12.5$meV ($V/E_F\approx 0.4$), (c) $V=18.75$meV ($V/E_F\approx 0.6$),
 and (d) $V=25$meV ($V/E_F\approx 0.8$).
The other parameters are taken relevant for InAs:
$\alpha=2\cdot 10^{-11}$eVm, $m^*=0.033 m_e$, $g^*=-12$, $E_F=30$meV.}
\label{ch1fig3}
\end{figure*}

\begin{figure*}
\includegraphics{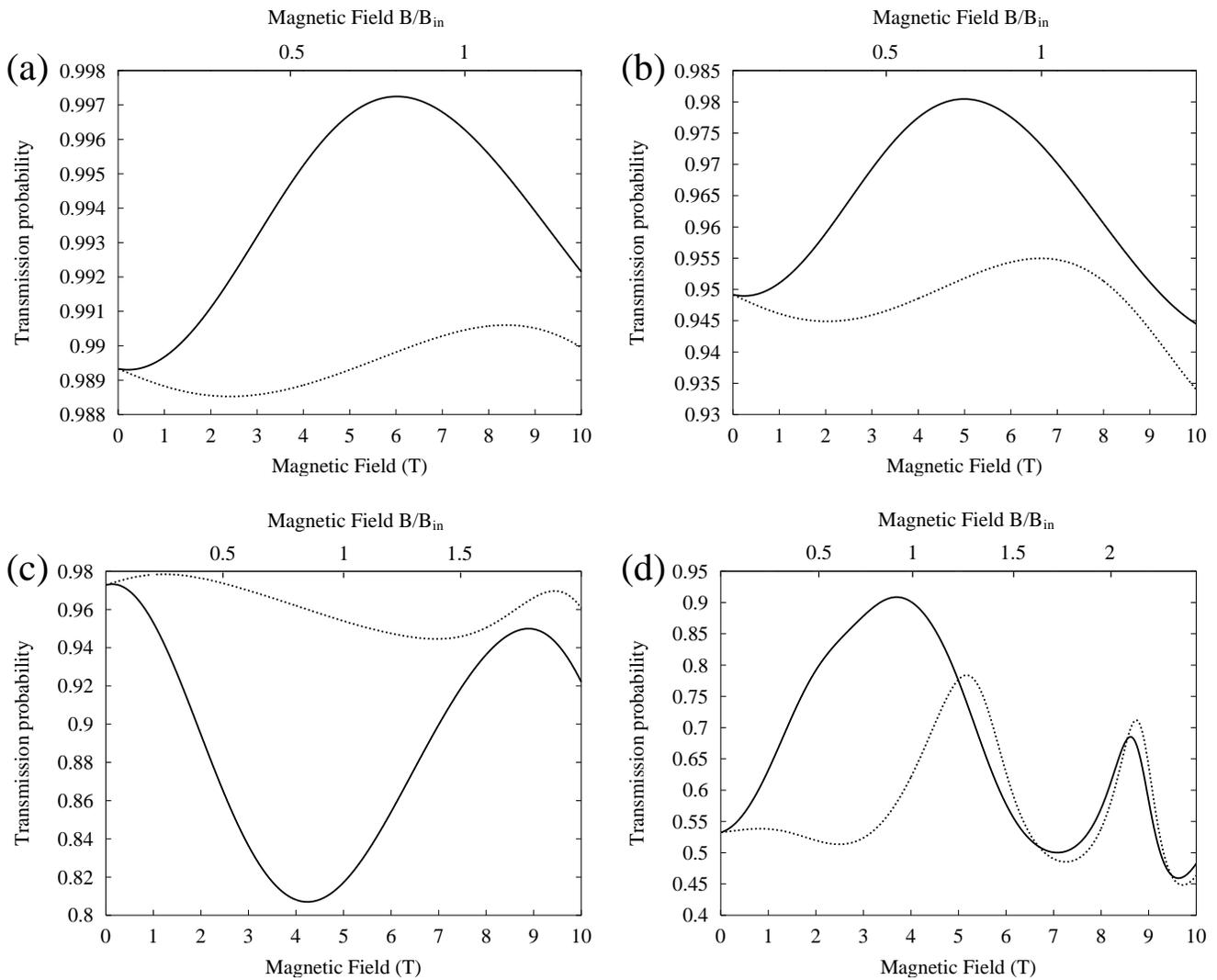}
\caption{Non-adiabatic regime $\hbar^2/\left(2m^* \alpha R\right) >> 1$.
The loop radius is taken ten times smaller than in
the previous figure (i. e. $R=5\cdot 10^{-6}$cm).
The other parameters for each panel are the same as for Fig.~\ref{ch1fig3}.}
\label{ch1fig4} 
\end{figure*}

\begin{figure*}
\includegraphics{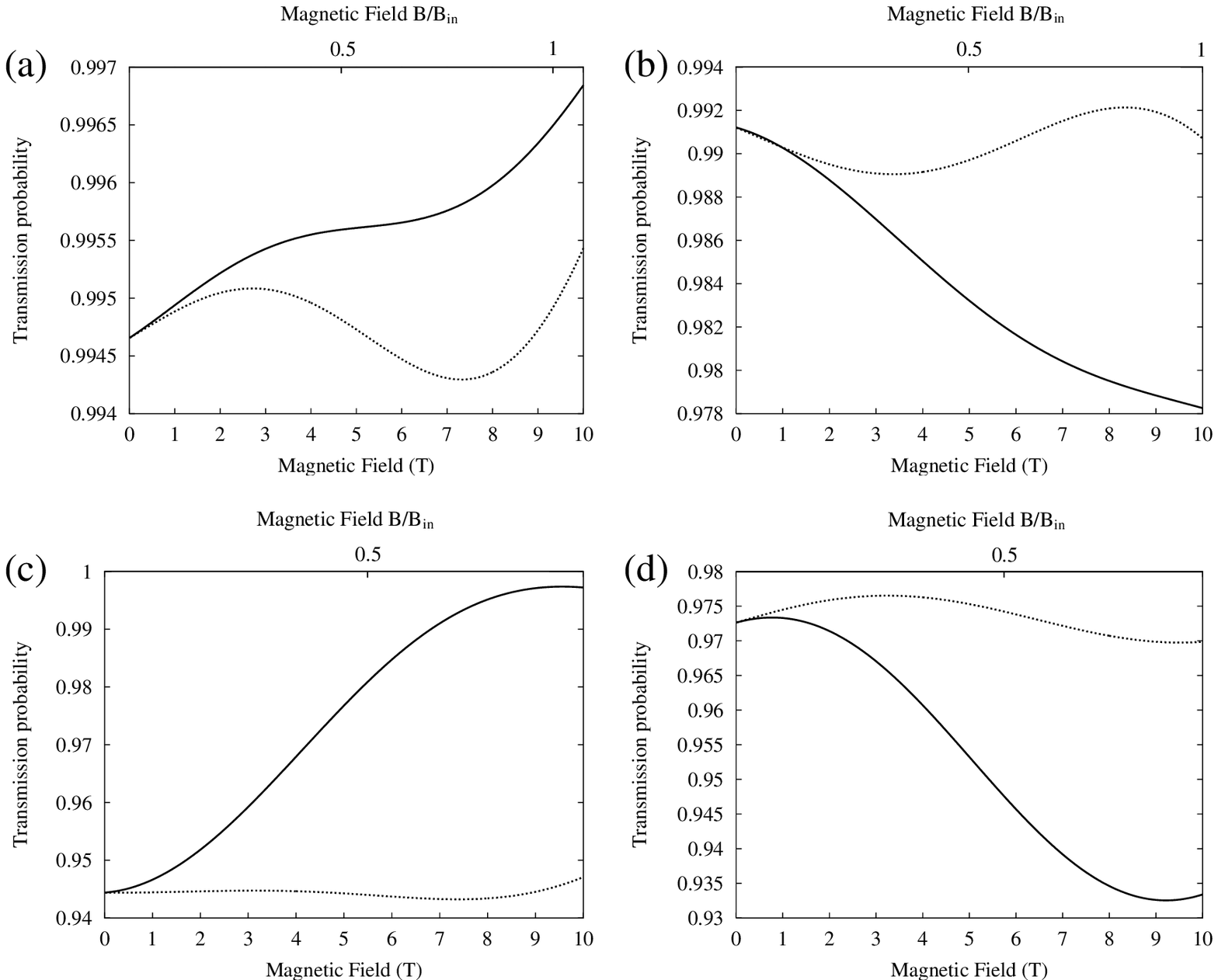}
\caption{Transmission probabilities for the loop 
in the non-adiabatic regime (solid lines) and corresponding straight wire of length
$L=2\pi R$ (dotted lines) versus external magnetic field.
The loop radius is $R=5\cdot 10^{-6}$cm.
The other parameters are the same as for Fig.~\ref{ch1fig3}, but
the height of the barrier is taken negative:
(a) $V=-6.25$meV,
(b) $V=-12.5$meV, (c) $V=-18.75$meV, and (d) $V=-25$meV.
Note, that the region of the magnetic fields,
where $B_z\gg B_\mathrm{in}$ (and where 
the difference between solid and dotted lines vanishes)
is shifted to very high values. }
\label{ch1fig6}
\end{figure*}

Since the main contribution to the current is given by the electrons at the Fermi level,
we consider the eigen states (\ref{input_psi+}) and (\ref{input_psi-})
at the fixed energy $E_F$.
Thus, the wave vectors $k^{\pm}$ in (\ref{input_psi+}) and (\ref{input_psi-}) are the Fermi ones,
and they satisfy the dispersion relations
\begin{equation}
\label{spectrum_wire}
E_F=\frac{\hbar^2 {k^\pm}^2}{2m^*}\pm
\sqrt{\alpha^2 {k^\pm}^2+\varepsilon_Z^2}.
\end{equation}
These equations give us four solutions with respect to $k$.
Each solution corresponds to the Fermi wave vector
with given chirality and spin indices. The absolute values of
the Fermi wave vectors with a given spin index for the left- and right-moving electrons
are equal in the straight channels.

The solutions (\ref{input_psi+}) and (\ref{input_psi-}) can be represented as
sums of incident and reflected waves.
The coefficients $A^\pm$ are the reflection amplitudes that have to be found imposing
the boundary conditions (\ref{conditions}).
For the output channel the reflection amplitudes are assumed to be zero,
and the corresponding spinors read
\begin{equation}
\label{out_psi+}
\Psi^+_\mathrm{out}(x)=\left(\begin{array}{l}
D^+\cos\gamma^+ {\mathrm e}^{i(k^+ + \frac{\Phi}{\Phi_0 R})x} \\
-iD^+\sin\gamma^+ {\mathrm e}^{i(k^+ + \frac{\Phi}{\Phi_0 R})x} \end{array} \right),
\end{equation}
\begin{equation}
\label{out_psi-}
\Psi^-_\mathrm{out}(x)=\left( \begin{array}{l}
-iD^-\sin\gamma^-  {\mathrm e}^{i(k^- + \frac{\Phi}{\Phi_0 R})x} \\
D^-\cos\gamma^- {\mathrm e}^{i(k^- + \frac{\Phi}{\Phi_0 R})x} \end{array} \right).
\end{equation}
Here $D^\pm$ are the transmission amplitudes.

The eigenfunctions of the Hamiltonian (\ref{loop_hamiltonian}) are of the form
\begin{eqnarray}
\nonumber
&& \Psi^+_\mathrm{loop}(\varphi) = 
{\mathrm e}^{i \frac{\Phi}{\Phi_0} \varphi}\times \\
&&
\left(
\begin{array}{l}
B^+ \cos\alpha^+ {\mathrm e}^{i(q^+_R - \frac{1}{2})\varphi}
+C^+\cos\beta^+ {\mathrm e}^{-i(\frac{1}{2} + q^+_L)\varphi}  \\
B^+ \sin\alpha^+ {\mathrm e}^{i(\frac{1}{2} + q^+_R)\varphi}
-C^+\sin\beta^+ {\mathrm e}^{-i(q^+_L - \frac{1}{2})\varphi}  \end{array} \right), 
\label{loop_psi+}
\end{eqnarray} 
\begin{eqnarray}
\nonumber
&& \Psi^-_\mathrm{loop}(\varphi)= {\mathrm e}^{i \frac{\Phi}{\Phi_0} \varphi}
\times \\
&&
\left(\begin{array}{l}
- B^- \sin\alpha^- {\mathrm e}^{i(q^-_R - \frac{1}{2})\varphi}
+C^-\sin\beta^- {\mathrm e}^{-i(\frac{1}{2} + q^-_L)\varphi}  \\
B^- \cos\alpha^- {\mathrm e}^{i(\frac{1}{2} + q^-_R)\varphi}
+C^-\cos\beta^- {\mathrm e}^{-i(q^-_L - \frac{1}{2})\varphi}  \end{array} \right),
\label{loop_psi-}
\end{eqnarray}
where
\begin{equation}
\label{alpha}
\tan\alpha^{\pm}= \frac{{\varepsilon_0 q^\pm_R-\varepsilon_Z}}{{q^{\pm}_R}\,{\alpha}/R}  +
\sqrt{1 + \left(\frac{\varepsilon_Z-\varepsilon_0 q^\pm_R}{q^{\pm}_R \,\alpha /R}\right)^2},
\end{equation}
\begin{equation}
\label{beta}
\tan\beta^{\pm}= - \frac{{\varepsilon_0 q^{\pm}_L + \varepsilon_Z}}{{q^{\pm}_L}\,{\alpha}/R}  +
\sqrt{1 + \left(\frac{\varepsilon_Z+\varepsilon_0 q^\pm_L}{{q^{\pm}_L}\,{\alpha}/R}\right)^2},
\end{equation}
and $q^{\pm}_{R,L}$ are the Fermi angular momenta in the curved part of the wire that are found from the
conditions which explicitly include the height of the barrier $V$
\begin{equation}
\label{qR_loop}
E_F = V+\frac{{\varepsilon_0}}{4} + \varepsilon_0{q_R^\pm}^2
\pm\sqrt{\left(\frac{{{q_R^\pm}}\,{{\alpha}}}{R}\right)^2 + {\left({q_R^\pm}\,{\varepsilon_0}
 -  {\varepsilon_Z} \right) }^2},
\end{equation}
\begin{equation}
\label{qL_loop}
E_F = V+\frac{{\varepsilon_0}}{4} + \varepsilon_0{q_L^\pm}^2
\pm\sqrt{\left(\frac{{{q_L^\pm}}\,{{\alpha}}}{R}\right)^2 + {\left({q_L^\pm}\,{\varepsilon_0}
 +  {\varepsilon_Z} \right) }^2}.
\end{equation}
It is interesting to note, that Fermi angular momenta for electrons with
opposite chiralities are not equal to each other ($q^\pm_L \neq q^\pm_R$). This effect stems
from the particular geometry of the system. Indeed, as soon as we 
assume $R\rightarrow\infty$ the relations (\ref{qR_loop}) and (\ref{qL_loop}) both
become equal to (\ref{spectrum_wire}), where $k^\pm_R=k^\pm_L$.
(The angular momenta must be substituted by their
linear analogue, i. e.  $k^\pm_R=q^\pm_R/R$, $k^\pm_L=q^\pm_L/R$.)
Thus, the chiral asymmetry of Fermi angular momenta is essentially of geometrical
origin.

The imposing of the boundary conditions (\ref{conditions}) on the wave functions (\ref{input_psi+}),
(\ref{input_psi-}), (\ref{out_psi+}) -- (\ref{loop_psi-}) gives us a system of eight equations.
That system definitely has an analytical solution with respect to
$A^\pm$, $B^\pm$, $C^\pm$ and $D^\pm$. However, the formulae for 
the amplitudes are extremely cumbersome. Therefore, we do not adduce them here.

At this point  it is pertinent to turn to the probability
current density calculations.
(In the following we call the probability
current density just current density.)
The conventional formula for current density \cite{landau1958}
is derived for the Hamiltonian where the
spin and orbital degrees of freedom are separable.
This is not the case in presence of spin-orbit interactions.
The correct formula for current density is a bit more complicated and reads
\begin{equation}
\label{main_current}
\mathbf{j}=\frac{\hbar}{2m^*}\left(\Psi_1\hat{k}_x^* \Psi_1^*
+ \Psi_1^* \hat{k}_x \Psi_1 +
\Psi_2\hat{k}_x^* \Psi_2^* + \Psi_2^* \hat{k}_x \Psi_2 \right)-
$$
$$
-\frac{i\alpha}{\hbar}\left(\Psi_1\Psi_2^*-\Psi_1^*\Psi_2\right),
\end{equation}
where $\Psi_1$ and $\Psi_2$ are two components of a given spinor.

Using the general relation (\ref{main_current}) one can easily find the input, reflected and
transmitted current densities for our particular system.
Note, that each current density is given as a sum of its two spin components 
$j=j^{+}+j^{-}$, and each component can be found using the following
formulas
\begin{equation}
\label{j_in}
j^\pm_\mathrm{in} = \frac{\hbar}{m^*}\left[k^{\pm}\pm\frac{\alpha m^*}{\hbar^2}\sin(2\gamma^\pm)\right],
\end{equation}
\begin{equation}
\label{j_refl}
j^\pm_\mathrm{refl} =-\frac{\hbar}{m^*} |A^\pm|^2\left[k^{\pm}\pm\frac{\alpha m^*}
{\hbar^2}\sin(2\gamma^\pm)\right],
\end{equation}
\begin{equation}
\label{j_out}
j^\pm_\mathrm{out} =\frac{\hbar}{m^*} |D^\pm|^2\left[k^{\pm}\pm\frac{\alpha m^*}
{\hbar^2}\sin(2\gamma^\pm)\right].
\end{equation}

\section{Results and discussion}

\label{sec3}

Now, we have everything ready to study the propagation of the initial states given
by (\ref{input_psi+}) and (\ref{input_psi-}) through the loop.
We define the transmission probability as
\begin{equation}
T=\frac{j_\mathrm{out}}{j_\mathrm{in}},
\end{equation}
while the reflection one reads 
\begin{equation}
R=\frac{j_\mathrm{refl}}{j_\mathrm{in}}.
\end{equation}

The plots of the transmission probability as a function of the external magnetic field
are shown in Figs.~\ref{ch1fig3}--\ref{ch1fig6} (solid lines) for 
different radii of the loop and barrier heights.
The additional dotted lines correspond to the
transmission probabilities through the wire of length $L=2\pi R$
separated from the input and output channels by barriers of the same height as the loop
is separated from its leads.
Therefore, it is only the curvature of the electron path
that differs for the solid and dashed lines.
The dependencies in Figs.~\ref{ch1fig3}--\ref{ch1fig6}
exhibit the following characteristic features.

First, the transmission probability oscillates as a function of the
external magnetic field $B_z$.
The oscillating factors appear in the transmission probability, because
of the interference between reflected and incident waves at the input and output of the loop.
It is well known, that the transmission probability for the quantum particle
propagating across a single rectangular potential barrier of length $L$
contains the oscillating factor $\sin(Lk)$, where $k$ is the wave vector of the 
particle \cite{landau1958}. 
Our case is a bit more complicated since we have two spin-split modes with
different wave vectors. Moreover, the absolute values of the Fermi angular momenta for the
left- and right-moving electrons with the same spin index differ from each other.
Therefore, we have many oscillating factors with different periods 
determined by $q^+_R$, $q^+_L$, $q^-_R$, $q^-_L$ and their combinations.
These angular momenta depend on the external magnetic field and, therefore,
the oscillations $T(B_z)$ occur.
We emphasize, that the fundamental 
origin of the oscillations depicted in Figs.~\ref{ch1fig3}--\ref{ch1fig6} is exactly the same
as in the simple single-mode model \cite{landau1958}.
In other words, our system is a kind of quantum interferometer with
the characteristic length $2\pi R$.

Here, we would like to emphasize the principal difference between
conventional interferometers based on the geometry
of a closed ring and Fabry-Perot-like system described above.
In both cases the interference pattern arises as a superposition
of the incident and reflected waves.
In the Aharonov-Bohm geometry reflected waves occur due to
the scattering of an incident wave on the contacts between upper/lower
arms of the ring and conducting leads, and the spin-orbit Berry phase manifestation
is usually found by comparison with the case of {\em negligible spin-orbit coupling},
where only ordinary Aharonov-Bohm effect take place.
In the loop geometry, the incident wave is reflected by the change of potential profile 
(see Fig.~\ref{ch1fig1a}),
and one can track the Berry phase manifestation by comparison
with the case of {\em a straight wire}.
The latter seems interesting from the theoretical point of view, since
the case of $R\rightarrow\infty$ is intractable in the model of a closed ring.

\begin{figure}
\includegraphics{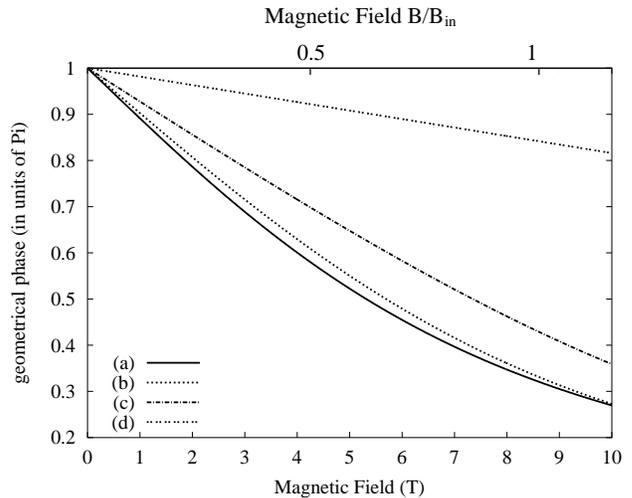}
\caption{The topological phase as a function 
of the external magnetic field at different loop radii:
(a) adiabatic approximation (\ref{berry}),
(b) $R=10^{-5}$cm, (c) $R=5\cdot 10^{-6}$cm, and (d) $R=10^{-6}$cm.
Magnetic field is given in Tesla (lower axis) as well as
in units of $B_\mathrm{in}$ (upper axis).
The barrier height $V$ is taken equal to zero,
and the other parameters are the same as for Fig.~\ref{ch1fig3}.}
\label{ch1berry}
\end{figure}

Second, there is a strong difference between transmission probabilities
for the loop and the straight wire at certain intermediate values of the magnetic field
(see Figs.~\ref{ch1fig3}--\ref{ch1fig6}),
while at higher values and at $B_z=0$ both curves just coincide.
This is a particular manifestation of the Berry phase that we explain in what
follows. First of all note, that the Berry phase is always  zero in the straight wire.
In contrast to that simple case,
an additional Berry phase dependent factor $\sin\phi_B$
occurs while an electron wave function propagates through the loop.
The Berry phase (\ref{berry}) is
negligible at $B_\mathrm{ext} \equiv B_z\gg B_{\mathrm{in}}$ and equal to $\pi$ at $B_z=0$
(see Fig.~\ref{ch1berry}).
Therefore, the factor $\sin\phi_B$ does not show up in these cases.
At certain intermediate values of $B_z$
the difference between straight wire and loop geometry is essential.
In  particular, at certain special values of the external magnetic field
the Berry phase is close to $\pi/2$ and the difference between transmission probabilities
for the loop and the straight wire is maximal.
We find it necessary to estimate such a magnetic field
using the quasi-classical formula (\ref{berry})
and assuming parameters relevant for InAs:
$\alpha=2\cdot 10^{-11}$eVm, $g^*=-12$, $k=10^6\mathrm{cm^{-1}}$.
Then, the Berry phase value $\pi/2$ corresponds to $B_z=|B_{\mathrm{in}}|/\sqrt{3}$
or, numerically, $\sim 3$T that is in good agreement with the plots
in Fig.~\ref{ch1fig3}.

The influence of the barrier height on the oscillations $T(B_z)$
is shown in Figs.~\ref{ch1fig3}--\ref{ch1fig6}.
First of all, one can easily see, that the transmission probability for the loop 
can also exceed its characteristic value for the straight wire.
Most importantly, however, the critical value of the magnetic fields
(where the difference between transmission probabilities for the loop
and straight wire is maximal) is very sensitive to the barrier height $V$.
This is explained in what follows. 

It is obvious, that the potential profile changes the Fermi momenta in the loop.
Since the Berry phase explicitly depends on the characteristic wave vector of the 
particle (\ref{berry}), we have a possibility to change the Berry phase by tuning
the potential profile. In detail, $B_\mathrm{in}$ is {\em proportional}
to the wave vector of the particle, whereas the Fermi momentum for
a given mode is larger for a deeper potential profile (i. e. for smaller 
or even negative $V$). Thus, the critical value of the external magnetic field
$B_z=B_\mathrm{in}/\sqrt{3}$ (which corresponds to $\phi_B=\pi/2$)
is shifted to the higher values when the electron bands are pulled down by $V$.
Moreover, at certain negative values of $V$ the Fermi wave vectors are so 
large, that the critical value $B_z=B_\mathrm{in}/\sqrt{3}$ exceeds $10$T, and, therefore,
the point, where Berry phase vanishes ($B_z\gg B_\mathrm{in}$)
leaves the reasonable  range of magnetic fields
depicted in Fig.~\ref{ch1fig6}.

Finally, let us make some important comments
on the role of the loop radius in the effect studied.
Indeed, further questions arise when we compare the plots
in Figs.~\ref{ch1fig3},\ref{ch1fig4}.
It is clearly seen, that the maximum of the difference
between transmission probabilities of the loop and the straight wire is shifted to higher magnetic fields.
However, the Berry phase does not depend itself on the radius of curvature.
Nevertheless, we can explain the effect if we remember,
that the formula (\ref{berry}) (and the Berry concept as well) is
valid only for the adiabatic motion. The latter means, that $\alpha m^* R/\hbar^2$ must be
lager than one, so that the electron spin precesses a few times while it is moving through
the loop. This is not the case depicted in Figs.~\ref{ch1fig4},
\ref{ch1fig6}, where $\alpha m^* R/\hbar^2\sim 0.5$
and the spin evolution is definitely not adiabatic.
Note, that our general approach is valid for both adiabatic and non-adiabatic cases,
because we use a direct solution of the Schr\"odinger equation.
Therefore, we are able to see the 
non-adiabaticity effects in Figs.~\ref{ch1fig4}, \ref{ch1fig6}.

The phase difference for particles moving in opposite directions
with the same spin index reads
\begin{eqnarray}
\nonumber
\phi_{\mathrm{top}}&=&\pi\left[1-(q^+_L - q^+_R)\right],\\
&=&\pi\left[1+(q^-_L - q^-_R)\right].
\label{top_phase}
\end{eqnarray}
Here, the index ``$\mathrm{top}$'' means ``topological'' since this phase is zero
in the straight wire.
Note, that exactly this phase difference is responsible for
the amplitude modulation of considered oscillations in the loop
as compared with the straight wire.
The Fermi angular momenta $q^\pm_{L,R}$ in Eq.~(\ref{top_phase})
are obviously radius dependent
[see Eqs.~(\ref{qR_loop}),(\ref{qL_loop})].
Thus, the geometric phase $\phi_{\mathrm{top}}$ 
is radius dependent as well (in the non-adiabatic regime).
As one can see from Fig.~\ref{ch1berry}, the topological phase is larger
than its adiabatic approximation (i. e. Berry phase)
for smaller radii of curvature.
Therefore, the characteristic magnetic fields,
which provide the maximal difference between the 
transmission probabilities of the loop and straight wire, are shifted
to their higher values for small loops.

At the end of the discussion,
we would like to clarify the relation between 
the topological phase $\phi_{\mathrm{top}}$ defined here and
phases described in earlier papers (i. e. Berry, Aharonov-Anandan, and
Aharonov-Casher).
As one can see from Fig.~\ref{ch1berry},
the topological (\ref{top_phase}) and Berry (\ref{berry}) phases give close results
in the adiabatic regime (large $R$). However,
$\phi_{\mathrm{top}}$ is {\em not} Aharonov-Anandan phase
$\phi_\mathrm{AA}$ as one may expect.
Ideed, the topological phase $\phi_{\mathrm{top}}$
is equal to $\pi$ at zero magnetic field for {\em any radius
of curvature}. (Berry phase is equal to $\pi$ here as well.) 
In contrast, Aharonov-Anandan and Aharonov-Casher phases are obviously
radius-dependent even at zero magnetic field \cite{PRB2004frustaglia,PRB2005molnar}
and read
\begin{equation}
\label{AA}
\phi_\mathrm{AA}=\pi\left(1-
\frac{1}{\sqrt{1+4m^{*2}\alpha^2 R^2/\hbar^4}}\right),
\end{equation}
\begin{equation}
\label{AC}
\phi_\mathrm{AC}=\pi\left(\sqrt{1+4m^{*2}\alpha^2 R^2/\hbar^4}-1\right).
\end{equation}
Thus, $\phi_{\mathrm{top}}$, $\phi_B$, $\phi_\mathrm{AA}$,
and $\phi_\mathrm{AC}$ are four different phases.

\section{Conclusions}

In conclusion, we have studied quantum transport in
a mesoscopic loop with Rashba coupling and Zeeman splitting.
Here, we have found that the Berry phase
gives a well pronounced effect in a form of a
deviation of the transmission probability from its value
for the straight wire of the same length $L=2\pi R$
at some specific external magnetic fields.
Moreover, we have investigated our system in the non-adiabatic regime
and found, that the characteristic magnetic fields, which provide the strong deviation,
are shifted to higher values. And finally, these specific values of the magnetic field
are very sensitive to the potential profile in the loop.

\begin{acknowledgments}
The authors acknowledge financial support by DFG from
Graduiertenkolleg ``Physik nanostrukturierter Festk\"orper'' (M.T.)
and Sonderforschungsbereich 508 (A. C.).
\end{acknowledgments}

\bibliography{refer.bib}

\end{document}